# Contrast modes in a 3D ion transmission approach at keV energies

R. Holeňák[1*], S. Lohmann[1] and D. Primetzhofer[1]

[1]Department of Physics and Astronomy, Uppsala University, Box 516, 751 20 Uppsala, Sweden

**Abstract**

We present options for visualizing contrast maps in 3D ion transmission experiments. Simultaneous measurement of angular distributions and flight time of ions transmitted through self-supporting, single-crystalline silicon foils allows for mapping of intensity and different energy loss moments. The transmitted projectiles were detected mainly for random beam-sample orientation using pulsed beams of He ions and protons with incident energies 50 and 200 keV. Differences in contrast, observed when varying the projectile type and energy, can be attributed to sample nuclear and electronic structure and bear witness to impact parameter dependent energy loss processes. Our results provide a base for interpretation of data obtained in prospective transmission studies when for example using a helium ion microscope.

Corresponding author: radek.holenak@physics.uu.se

## 1. Introduction

Thin self-supporting targets permit analysis of their composition and structure in transmission, gaining additional information in comparison to other spectrometric and spectroscopic approaches. Various probes, based on different particles, have been employed. Scanning transmission ion microscopy using MeV primary ions (STIM) is analogous to scanning transmission electron microscopy and several modes for contrast imaging were investigated with the emergence of microbeams in the 80s [1–3]. However, for MeV ions the beam spot size is limited, typically to μm [4,5] with a few studies with nanometre beams existing [6]. Even at much lower keV energies, the short wavelength of helium ions still results in a reduced diffraction limit for helium ions compared to electrons. Recently, a new imaging technique, the helium ion microscope (HIM) based on keV ion beams has been introduced. It offers high spatial resolution (<0.5 nm), large depth of field, and the ability to image nonconductive samples [7–9]. Along with secondary electron detection, it has so far been shown that contrast can be achieved by detecting primary and secondary ions scattered or sputtered from the sample surface [10]. For thin samples, however, a majority of incident helium ions is transmitted through the sample, i.e. >90 % of the helium ions scatter in forward direction, whereas only ~5 % or less are scattered in an angle larger than 90°. Due to this property, detection of ions in transmission would enable imaging of thin samples with low fluence drastically reducing the amount of sample damage to bulk and surface by introducing defects from small and large angle nuclear scattering and due to effective sputtering by keV ions, respectively [11].

Contrast in any transmission microscope using ions is related to either lower intensity as particles are scattered away or the energy loss that the projectile suffers during the passage through the sample. In the dark field mode Woelh *et al.* showed the advantages of detecting transmitted ions for quantitative imaging over conventional transmission electron microscopy (TEM) [12]. In the broad-beam setup introduced by Mousley *et al.* the intensity distribution patterns could also be related to the surface morphology [13]. Energy loss contrast originates either from different stopping, different thickness, or different scattering, i.e. different angular distribution for a given energy loss. The most commonly observed contrast comes from the variation of sample thickness and materials. However, even for a monoatomic compound of constant thickness, the intensity and the energy loss can show a dependence on the sample orientation if the target is crystalline. For MeV ions, the energy loss is known to show a strong impact parameter dependence, due to impact parameter dependence of inner-shell excitations [14–17] resulting in reduced loss along low-index crystal directions. Also, the intensity distribution of transmitted He projectiles is affected by these channelling and blocking effects, which can enhance or reduce the particle fluence in specific directions [18]. In a recent work Lohmann *et al.* showed that, even at energies typically employed in the HIM, the energy loss experienced by primary ions can depend strongly on the ion trajectories, due to a pronounced impact parameter dependence of the mean charge state [19,20].

Particle optics and dimension constraints in transmission approaches using the HIM are still a challenge for direct mapping of energy loss of transmitted ions. Wang *et al.* have constructed a transmission channelling setup for a He ion microscope [21], and recently, Kavanagh *et al.* used this setup to acquire transmission images showing an intensity contrast pattern [22].

In this work, we show how sample structure affects intensity as well as energy loss and its higher moments for both helium ions and protons transmitted through thin, self-supporting, single-crystalline silicon foils. For this aim, we simultaneously record angular distributions and particle energy and present different ways to obtain and visualize contrast, which is considered to be useful for potential future investigations using helium ion microscopes in transmission mode.

## 2. Experiment and data analysis

Experiments were performed with the time-of-flight medium energy ion scattering (TOF-MEIS) setup at Uppsala University [23,24]. Based on a Danfysik implanter, a chopped beam of atomic ions was provided to the experimental chamber, while maintaining a beam spot size smaller than 1x1 mm$^2$ and beam angular divergence better than 0.056°. The current impinging on the sample is 2-3 fA with a beam pulse width of 1 ns to 2 ns and a repetition rate of 31.25 kHz. In the present study, we focus on protons and $^4$He$^+$ primary ions as they feature maximum relevance for the HIM. The initial ion energies range between 50 and 200 keV. Projectiles are detected after transmission through 200 nm thin, self-supporting Si (100) foils. The angular distribution of transmitted particles is recorded with a large, position-sensitive microchannel plate detector with diameter 120 mm from RoentDek [25]. The detector covers a solid angle of 0.13 sr, and the position of particles is determined with help of two perpendicular delay lines. The ion energy, after transmission through the sample, is measured via its flight time.

The acquisition software COBOLDPC (Computer Based Online offline Listmode Dataanalyser) stores the data in a list mode file format [25]. Every time event can consist of up to five incidents (hits or noise) while every hit is assigned a XY position and flight time [26]. The assignment of the position is made from the time difference of the signals appearing at the two ends of each delay-line anode. When loading the file to the COBOLD interface, data is processed by an inner algorithm with conditions defined by the user. Custom changes in the source code allow us to extract an ASCII file containing a list of events holding position coordinates and the associated energy of the particle derived from its time-of-flight value. Multiple hits are rare and are, for the sake of simplicity of coding, excluded.

By binning the data in three dimensions, i.e. in energy and two spatial dimensions, statistical manipulation becomes accessible. Every spatial bin contains thus an energy distribution histogram of particles arriving at the same detector position. These histograms permit extraction of statistical values such as the most probable energy or the mean energy as well as its standard deviation (SD). These variables can be assigned to their respective position and plotted as 3D maps as the one shown in Figure 1. Spatial binning is for statistical reasons chosen as 0.5 mm, corresponding to an angular resolution ~0.1°. Spatial intensity distribution maps are normalized to the maximum value. Energy is binned with 1 keV steps, with regard to the fact that the detector energy resolution for all used projectiles should be better than 1 keV.

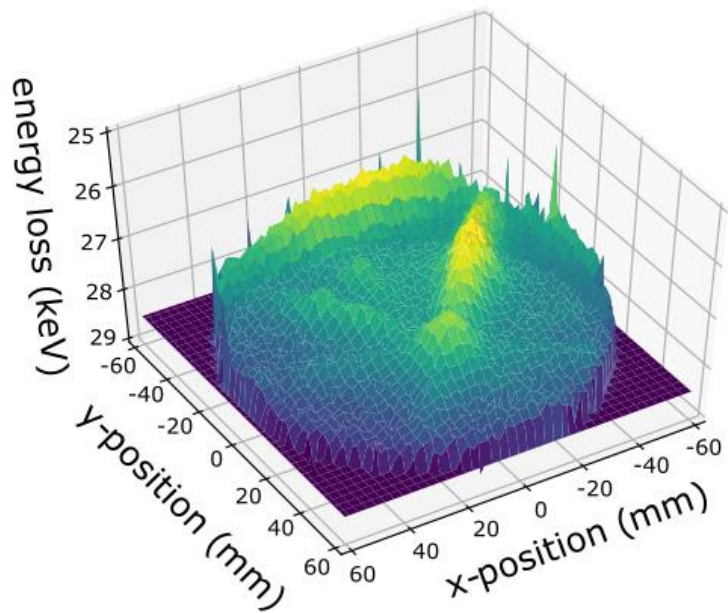

*Figure 1: 3D surface map of the mean energy loss for 50 keV He transmitted through a 200 nm single-crystalline silicon foil in a pseudo-random direction, i.e. rotated with respect to the [100] crystal axis by θx = 7.5° and θy = 12° around the x and the y axis, respectively.*

## 3. Results

For the primary beam oriented perpendicular towards the Si (100) surface, channelling allows the majority of the ions to propagate in a straight direction encountering no large-angle collisions. Intensity mapping on the detector features an intense peak in the centre with lower intensity surrounding it, which is formed by ions that escaped the channel and experienced scattering by larger angles than the critical angle for channelling [27] (see Figure 2). Some of these trajectories become a subject to blocking, which for 200 keV He, results in observable weak streaks along low-index planar channels. The same effect is taking place for 50 keV He, however, due to the wider angular distribution, it is not visible within the angle covered by the detector.

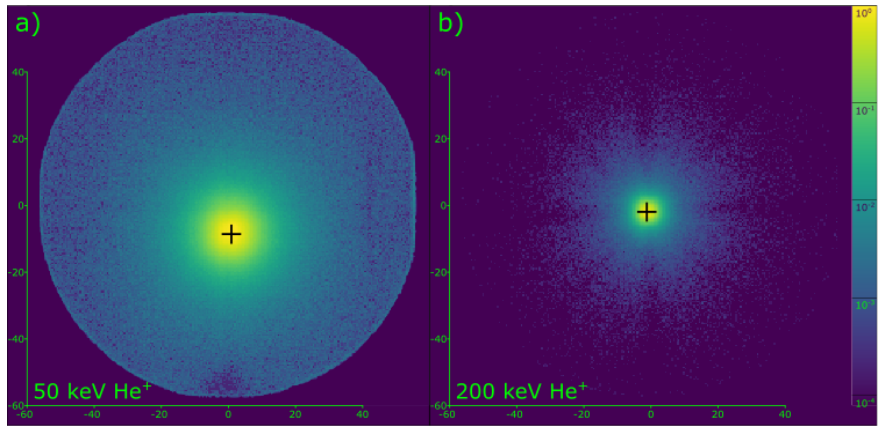

*Figure 2: Intensity distribution of transmitted a) 50 keV and b) 200 keV He entering the crystal in channelling orientation along [100] axis. Due to decreased scattering probabilities at higher projectile energy the combined channelling and blocking features can be observed within the detector opening angles only at 200 keV. Both maps are normalized to the maximum intensity. The initial beam position is marked with the black cross.*

More prominent, distinct features start to appear when the alignment of low-index crystal directions and beam is broken. Figure 3 holds three different contrast maps extracted from raw data for 50 keV He transmitted through 200 nm Si. The initial direction of the beam is pointing towards the centre of the MPC detector and is indicated by the white cross. The degree of alignment between beam and crystal axis away from the [100] channelling geometry is $θ_x$ = 7.5° and $θ_y$ = 12° corresponding to a rotation around the x and the y axis, respectively. Under such a geometry the beam is not aligned with any major crystal axis or plane, this condition is often referred to as (pseudo-) random. Animation 1 (see supplementary material) illustrates how the energy and intensity (as an inset) of the particles are affected by changing the orientation of the sample.

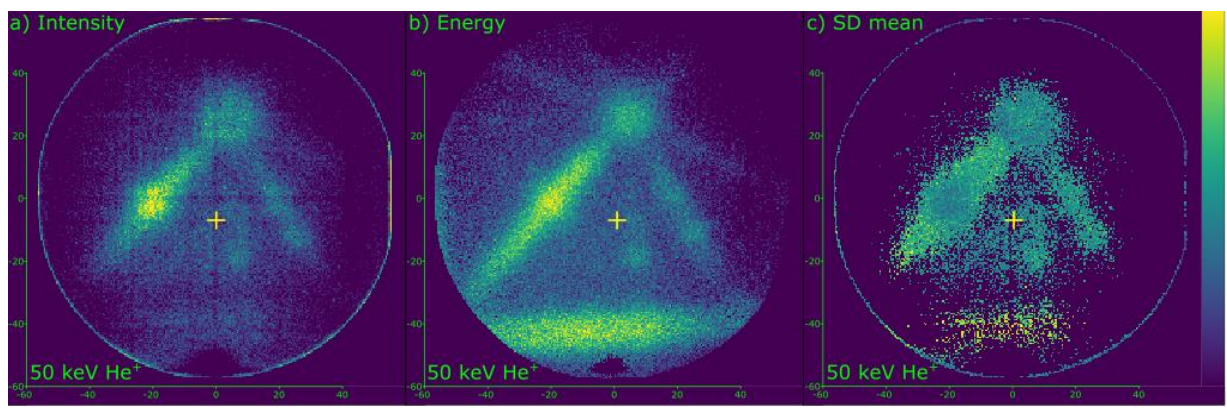

*Figure 3: Three different ways to visualize the contrast from acquired data: a) Direct intensity distribution on the MCP detector visualizing ion channelling through low-index axes and planes, b) Mean energy calculated in binned data and assigned to the respective position. Projectiles exiting the channels arrive at the detector with higher energies. c) Calculated standard deviation of mean representing the broadening of energy distribution along the planes. Initial beam direction is marked with the yellow cross*

Figure 3 a) straightforwardly shows the spatial intensity distribution of transmitted ions in the above-mentioned geometry. Only minor intensity is observed in the original beam direction. Instead, the observed intensity maxima, i.e. maximum contrast originates from trajectories of transmitted ions governed by channelling in a certain crystal direction. Specifically, maximum intensity is observed along the {110} plane, and the intensity in the [114] axial channel significantly exceeds the intensity in the original beam direction.

As described in the previous section, it is possible to calculate the mean energy of the ions detected for the same dataset. This representation is equivalent to plotting the energy loss experienced. Figure 3 b) shows the resulting spatial energy distribution of detected projectiles, i.e. virtually the same data as in Figure 1 but in 2D. With respect to the intensity distribution, the energy map gives a clearly different, and more pronounced contrast. It is readily visible that the ions lose significantly less energy along channelled trajectories compared to random trajectories. Also, more channels are visible and structures observed show a higher resolution.

A combination of the representations in Figures 3 a) and 3 b) is possible in an animation, even more clearly visualizing the observations. By slicing the spatial intensity distribution in the third dimension, i.e. energy, a single contrast map can be assigned to each energy loss value along the projectile's energy distribution (see the supplementary animation). Based on these differences in data presentation, the origin of the contrast in the {001} plane at the bottom side of the triangle in Figure 3 b) can be explained as a low number of projectiles arriving at the detector with relatively high energy after being channelled. Their total number is, however, statistically insignificant and thus highly suppressed in the intensity contrast in Figure 3 a).

To visualize energy distribution effects, the standard deviation of the mean was also calculated in every spatial bin and plotted in Figure 3 c). Distinct features can indeed be partially seen as the broadening of the signal in comparison to the intensity and energy loss mapping. For statistical reason, bins with low counts were blanked. The minimum value was set to the value, for which the calculated SD contrast disappeared from the detector's blind region at the bottom edge of the detector.

The contrasts illustrated in the previous examples are, however, not all of general character but can show a strong dependence on ion species as well as the employed ion energy. Figure 4 a) shows the behaviour of 200 keV He in identical geometry as employed for the measurements visualized in Figure 3. In comparison to the previous example, ions are propagated much more efficiently along trajectories almost parallel to the initial beam arriving at the centre of the detector. 50 keV protons in Figure 4 c) show a somewhat similar behaviour, i.e. the maximum intensity is still close to the original beam direction. They feature, however, a much broader overall distribution of ions scattered out of straight direction. Different from He at 50 keV, intensity is suppressed in channelling directions, as long as they are not close to the incident direction. Indeed, three minor channels show an increased intensity, whereas the {110} plane shows reduced intensity. Both energy maps in Figure 3 b) and Figure 4 b) for 50 keV and 200 keV He, respectively, exhibit an energy contrast between channelling and random direction. However, for 200 keV, the contrast is suppressed and only clearly visible for minor planes close to the initial beam direction. This observation is consistent with the observation of a maximum in the energy loss ratio for channelled and random trajectories as shown by data from Lohmann *et al.* and literature [19,28]. In contrast, data for 50 keV protons does not show any clear contrast in the energy distribution, which goes beyond a weak contrast that originates from reduced sample thickness towards the [100] direction, i.e. the lower left corner of the detector.

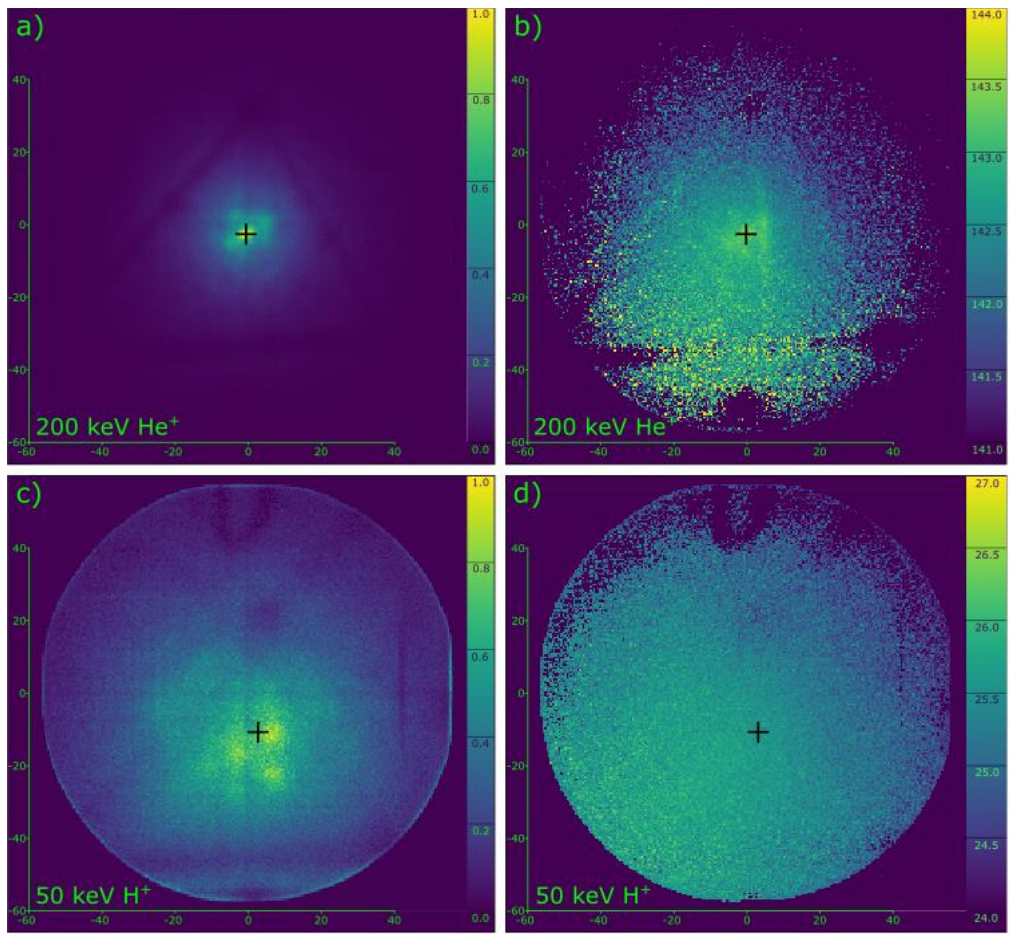

*Figure 4: Comparison between intensity and energy contrast for He ions and protons at 200 keV and 50 keV, respectively. Intensity contrast in a) and c) indicates that ions are propagated along the initial direction while entering higher index channels. Contrast in energy loss shows a projectile dependency when showing contrast only for helium. Intensity maps a) and c) are normalized to maximum intensity value in the map. Beam position is marked with a black cross.*

## 4. Discussion

Contrast in intensity reveals information about the crystal structure and about the scattering processes experienced by the projectiles. As shown for 50 keV He, even though the beam is aimed towards the centre of the detector, the maximum intensity can be observed in close-by channels and planes. We observe a clear dependence of the intensity distribution on both initial energy and projectile charge. As the scattering cross section increases for lower energy and increasing $Z_1$, steering of the projectiles along crystallographic axes and planes becomes more likely. This can be observed for 50 keV He where the projectiles are exiting the crystal from mostly low-index channels. At 200 keV, however, these channels become less accessible due to significantly decreasing critical angles which eventually results in predominant blocking from the same crystallographic features. Nevertheless, a significant number of ions are entering higher index channels located close to the beam due to the smaller, and thus more probable, deflection angles necessary (the bright spots around the central position). Also, 50 keV protons experience less scattering than helium at the same energy, and tend to propagate in a straight direction. The channelling into higher index axes is found more pronounced than in the case of 200 keV He. This behaviour can be well understood qualitatively, as both the Rutherford cross section but also the here more applicable screened potential predicts the scattering cross section for 50 keV protons to be found between those for 200 keV and 50 keV He ions. For a more quantitative discussion, one could compare the half-widths of the angular distributions with predictions for the channelling half angle [27] or perform Monte-Carlo-simulations. Such angular distributions can, however, not be obtained from the present samples, but would require amorphous foils.

Contrast maps of the mean energy of arriving particles can be linked to the energy loss processes taking place along the ion trajectories. In comparison to the intensity maps, this kind of contrast can be much more pronounced. Both low-index and high-index planes in Figures 3b) and 4b) are found narrower, and often better visible. The almost complete absence of visible contrast in the energy map for protons in Figure 4 d) indicates a strong projectile dependency of this kind of contrast. As for the detector opening angle, kinematic losses are negligible, even in multiple scattering by the maximum angle (half-opening angle of the detector) the overall energy loss can be attributed directly to the average electronic excitations per unit path length, i.e. electronic stopping. Dependence of the projectile energy loss along different crystal directions has been demonstrated for MeV projectiles by many experiments and various DFT calculations [29,30]. For the keV range are, to our knowledge, available simulations and data very scarce. Lohmann *et. al* have shown that the energy loss of channelled protons at employed energy reaches about 0.95 of the random energy loss. Such a low difference thus explains the lack of significant energy contrast in the map in Figure 4 d). For helium ions, the reverse trend was observed when the channelled energy loss only reaches 0.82 of the random one at 50 keV, but increases to about 0.9 at 200 keV [19]. The gradual contrast observed diagonally towards the left corner in the Figure 3 d) is a result of increased length of the particle trajectory through the sample, which is expected to be max 10 % over the whole angular span of the detector.

In light of the discussion above, one can attempt to carefully interpret the map of the standard deviation of the mean. At first, however, one should discuss a statistical issue in the calculation of SD that needs to be handled before visualizing the data. Detector areas with low count intensity, in combination with background noise, i.e. white noise at the channel plate detector, can result in huge standard deviation thus saturating the final contrast. Therefore, a threshold of a minimum of 100 counts per pixel, selected as described above, was applied as a condition for visualization. This standard deviation can be understood as a form of energy straggling of ions with possibly different trajectories inside the crystal, but similar final angular deflection. One could expect such contrast to be visible at the edges of the axial and planar channels i.e. in the transition regime from channelling to random geometry, with qualitatively very different trajectories (analogy to the skimming effect described for backscattering [31]), resulting in signal broadening. There are, however, other contributions from e.g. higher angle scattering and dechannelling that complicate the straightforward interpretation of the observed contrast. As a result, the observed distributions are far from what is commonly considered as energy loss straggling, i.e. close to Gaussian distribution of the observed particle energies. In turn, the specific sample structure will have a huge influence on how data can be interpreted. As such, an interpretation of straggling maps, due to the huge amount of mixed information they contain, is expected to be only meaningful based on complete pictures from both intensity and energy distributions.

## 5. Summary

We have employed time and position resolved ion transmission experiments to shed light on the possibilities of contrast imaging in scanning transmission ion microscopes. Different primary ions and energies were employed. Three complementary ways of contrast mapping of intensity, mean energy and standard deviation obtained from the same initial datasets are shown and discussed, each exhibiting distinct features. The data provided visualizes how images recorded using a helium ion microscopy employed in transmission geometry can yield information on different aspects of sample structure.

**Acknowledgments**

Accelerator operation is supported by the Swedish Research Council VR-RFI (Contracts No. 821-2012-5144 and No. 2017-00646_9) and the Swedish Foundation for Strategic Research (Contract No. RIF14-0053).

**Supplementary material**

*Animation 1.tif*: Change in energy and intensity (as an inset) contrast with respect to sample orientation. Channelling along the low-index axes and planes can be observed as the beam moves from the initial [100] direction towards a pseudo-random orientation. The projected direction of the primary beam is marked with a cross.

*Animation 2.tif:* Spatial distribution of projectiles which experienced the same energy loss during transmission. The most energetic ions exit the crystal in channel directions whereas the slower ions are distributed homogenously on the detector, indicating higher energy loss per unit path length along their trajectories.